\documentclass[manuscript]{aastex}
\usepackage{epsfig}
\usepackage{amsmath}
\usepackage{soul}
\usepackage{hyperref}
\hypersetup{
    unicode=false,          
    pdffitwindow=false,     
    pdfstartview={FitH},    
    pdftitle={My title},    
    pdfauthor={Author},     
    pdfsubject={Subject},   
    pdfcreator={Creator},   
    pdfproducer={Producer}, 
    pdfkeywords={keyword1, key2, key3}, 
    pdfnewwindow=true,      
    colorlinks=false,       
    linkcolor=red,          
    linkbordercolor={1 0 1},
    citebordercolor={0 1 0},
}
\begin{document}

\title{The Fundamental Vibrational Frequencies and Spectroscopic Constants of
the Dicyanoamine Anion, NCNCN$^-$ (C$_2$N$_3$$^-$): Quantum Chemical Analysis for Astrophysical and Planetary Environments}

\date{\today}
\author{David Dubois}
\affil{MS 245-6 NASA Ames Research Center, Space Science \& Astrobiology Division, Astrophysics Branch, Moffett Field, California 94035,
U.S.A.}
\affil{Bay Area Environmental Research Institute, Moffett Field, California 94035, U.S.A.}
\author{Ella Sciamma-O'Brien}
\affil{MS 245-6 NASA Ames Research Center, Space Science \& Astrobiology Division, Astrophysics Branch, Moffett Field, California 94035, U.S.A.}
\author{Ryan C. Fortenberry}
\email{r410@olemiss.edu}
\affil{Department of Chemistry \& Biochemistry, University of
Mississippi, University, Mississippi 38677-1848, U.S.A.}
\begin{abstract}

Detecting anions in space has relied on a strong collaboration between theoretical and laboratory analyses to measure rotational spectra and spectroscopic constants to high accuracy. The advent of improved quantum chemical tools operating at higher accuracy and reduced computational cost is a crucial solution for the fundamental characterization of astrophysically-relevant anions and their detection in the interstellar medium and planetary atmospheres. In this context, we have turned towards the quantum chemical analysis of the penta-atomic dicyanoamine anion NCNCN$^-$ (C$_2$N$_3$$^-$), a structurally bent and polar compound. We have performed high-level coupled cluster theory quartic force field (QFF) computations of C$_2$N$_3$$^-$ satisfying both computational cost and accuracy conditions. We provide for the first time accurate spectroscopic constants and vibrational frequencies for this ion.
In addition to exhibiting various Fermi resonances, C$_2$N$_3$$^-$ displays a bright $\nu _2$ (2130.9 cm$^{-1}$) and a less intense $\nu _1$ (2190.7 cm$^{-1}$) fundamental vibrational frequency, making for strong markers for future infrared observations $<5 \mu m$. We have also determined near-IR overtone and combination bands of the bright fundamentals for which the $2\nu_2$ at 4312.1 cm$^{-1}$ (2.319 $\mu m$) is the best candidate. C$_2$N$_3$$^-$ could potentially exist and be detected in nitrogen-rich environments of the ISM such as IRC +10216 and other carbon-rich circumstellar envelopes, or in the atmosphere of Saturn's moon Titan, where advanced N-based reactions may lead to its formation.

\end{abstract}

\section{Introduction}

The chemical reservoir of negative ions in the universe encompasses a broad range of astrophysical and astrochemical environments: from cold interstellar clouds, circumstellar envelopes and carbon-rich Asymptotic Giant Branch (AGB) stars, to cometary comae and planetary atmospheres. The combined theoretical and experimental prospect for their spectroscopic characterization has resulted in unprecedented observations of key cosmic anions in the aforementioned environments (see review by \cite{Millar2017} and references therein).

Since the first anion observations in dense molecular clouds and circumstellar envelopes (CSE) in 2006 \citep{McCarthy2006}, their detection has remained difficult. The well-studied IRC +10216 CSE \citep[e.g.][]{Remijan2007} has nonetheless revealed the presence of C$_n$H$^-$ carbon-chain as well as C$_{n-1}$N$^-$ cyano molecular anions \citep[e.g.][]{Thaddeus2008} with $n = 2-6$. Among these, C$_8$H$^-$ \citep{Brunken2007} represents the largest carbon-chain anion detected in the interstellar medium (ISM).  Molecular anions containing multiple nitrogen atoms are still elusive as their experimental and theoretical spectroscopic constants are no yet available to guide observations.

The direct and indirect molecular study of C$_2$N$_3$$^-$ can be traced back to work carried out over the past two centuries, with a specific emphasis on the understanding of peptide and polypeptide synthesis. Historically, interest was drawn upon cyanamide CH$_2$N$_2$ as early as 1858, when cyanamide was shown to readily dimerize into dicyanamide C$_2$N$_3$H \citep{Beilstein1858}. It was not until 1922 that a first synthesis scheme for this molecule and a series of other H$_x$C$_y$N$_z$ compounds involving de-ammonated reactions were proposed, for which a C$_2$N$_3$ nucleus was identified \citep{Franklin1922}. These derivatives were further investigated by \cite{Steinman1966} and \cite{Steinman1967} in the study of dicyanamide as an intermediate in the synthesis of polypeptides such as diglycine in the primitive Earth environment. Their scheme also included the dicyanoamine anion NCNCN$^-$. Henceforth, C$_2$N$_3$H and C$_2$N$_3^-$ were determined to be important precursors in the prebiotic chemical evolution leading to the formation of peptides. \cite{Schimpl149} also investigated the formation of cyanamide under primitive Earth conditions. In addition, reactive and thermodynamic properties of NCNCN$^-$ in ionic liquids were investigated by \cite{Dahl2005, Jagoda-Cwiklik2007, Nichols2016}.

Recently, observations from the Cassini mission unveiled the presence of unidentified large anions with masses up to 13,800 \textit{u/q} at high altitude ($>$ 950 km) in the atmosphere of Titan, Saturn's largest moon \citep{Coates2007}.  Recent work has helped characterizing observed mass peaks and growth patterns in relation to ion N-based reactions \citep{Lavvas2013, Desai2017, Dubois2019}. Furthermore, C$_2$N$_3$$^-$ was reportedly identified as a potential precursor seed fragment in the polymeric growth of laboratory-produced atmospheric aerosols, analogs of Titan's atmospheric haze \citep{Carrasco2009, Somogyi2012}.\

While even available for purchase from many chemical companies as a solid salt
of sodium dicyanoamide, the gas phase vibrational and rotational spectra of
this molecule have yet to be conclusively provided.  Such spectroscopic data would be required for
any future observations of this molecule in planetary atmospheres, CSEs, or other astrophysical environments.  As such, the
work presented here relies upon proven quantum chemical tools for the predictions of these
values.  The established CcCR approach utilizes a quartic force field
(QFF, fourth-order Taylor series expansion of the internuclear Hamiltonian)
comprised of energy points defined from quantum chemical complete basis set
extrapolation (``C''), inclusion of core-electron correlation (``cC''), and
considerations for scalar relativity (``R'') \citep{Huang08, Huang09, Huang11,
Fortenberry11HOCO}.  This method has produced vibrational frequencies to within
1.0 cm$^{-1}$ and rotational constants within 10 MHz of gas phase experiment in
many cases \citep{Huang08, Huang09, Huang11, Zhao14, Fortenberry12HOCO+,
Huang13NNOH+, Fortenberry14C2H3+, Fortenberry15SiC2, Kitchens16, Bizzochi17,
Fortenberry17IJQC}.  Recently, the use of explicitly correlated electron
wavefunctions has promised an increase in quantum chemical accuracy at
significantly reduced computational cost.  Contemporary work has shown that the
coupled cluster singles, doubles, and perturbative triples [CCSD(T)] method
\citep{Rag89} within the F12 explicitly correlated formalism \citep{Adler07,
Knizia09, Kong11} is typically within 6.0 cm$^{-1}$ or better of the expected
CcCR results but with orders of magnitude reduction in computational time
\citep{Agbaglo19a, Agbaglo19b}.  As such, both highly-accurate approaches are employed here to produce the necessary vibrational and rotational spectroscopic
data for NCNCN$^-$ in order to aid in its potential classification in the
atmosphere of Titan, IRC +10 216, or other celestial objects.

\section{Computational Details}

\begin{figure}[h]
\includegraphics[width = 6.2 in]{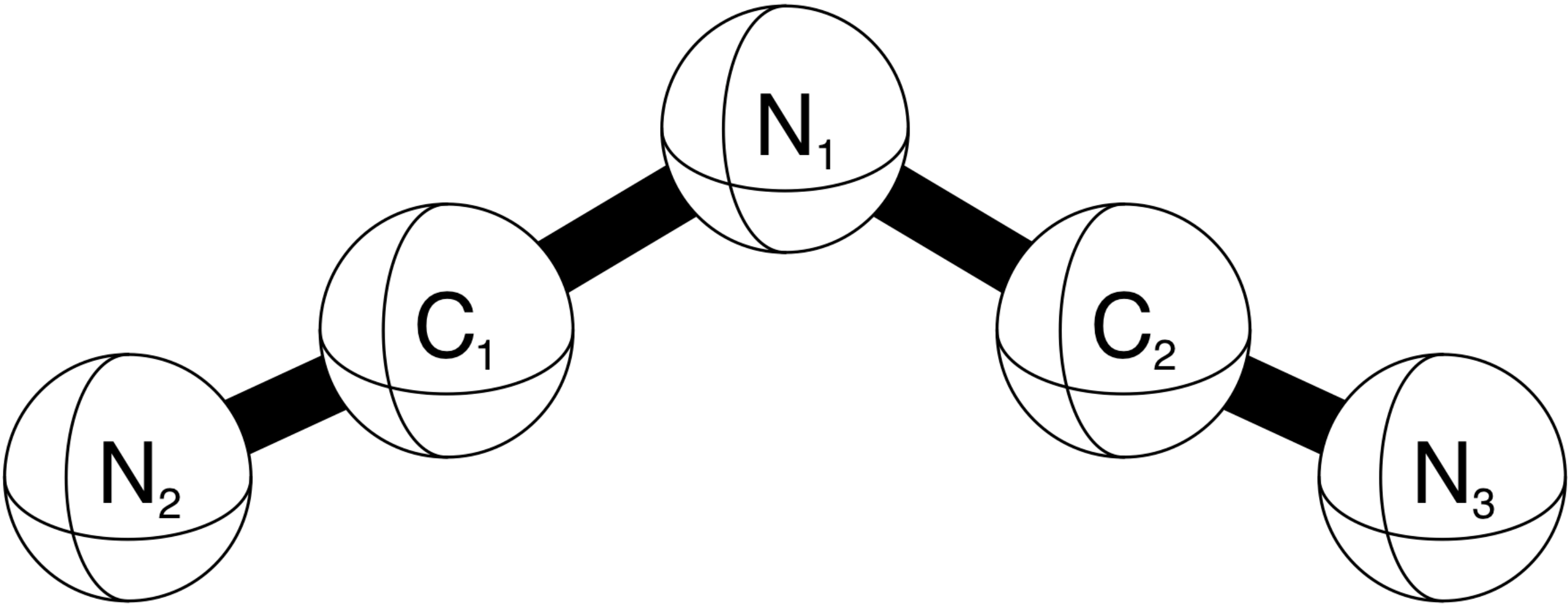}
\caption{The Equilibrium Structure of NCNCN$^-$.}
\label{fig}
\end{figure}

The two different approaches used in this study are detailed hereafter. The QFF procedure begins with a geometry optimization and construction of the reference geometry.  The CcCR geometry is defined from a CCSD(T)/aug-cc-pV5Z \citep{aug-cc-pVXZ, Dunning89} minimum further corrected for core correlation
from optimizations with the Martin-Taylor (MT) core correlating basis set
\citep{Martin94} both including and excluding the core electrons.  The
CCSD(T)-F12/cc-pVTZ-F12 (denoted as F12-TZ from here on) QFF, on the other hand, begins simply with
a an optimized minimum at this level of theory with no further corrections.
Regardless of the method, the QFF is defined from 1585 points based on
displacements of the reference geometry using the below $S_i$ symmetry-internal
coordinates defined by Equ. 1--9 and an efficient lazy-Cartesian algorithm to construct the
displacements \citep{Thackston18}.  The bond lengths are displaced by 0.005
\AA\ and the angles as well as the in-plane (LINX) and out-of-plane (LINY)
near-linear bends are displaced by 0.005 radians.  The atom numbering and symmetry-internal coordinates, $a_i$ and $b_i$, in this
$C_{2v}$ molecule are done according to Figure \ref{fig}.
\begin{align}
S_1(a_1) &= \mathrm{C_1}-\mathrm{C_2}\\
S_2(a_1) &= \frac {1}{\sqrt{2}}[(\mathrm{N_1}-\mathrm{C_1})+(\mathrm{N_1}-\mathrm{C_2})]\\
S_3(a_1) &= \frac {1}{\sqrt{2}}[(\mathrm{C_1}-\mathrm{N_2})+(\mathrm{C_2}-\mathrm{N_3})]\\
S_4(a_1) &= \frac {1}{\sqrt{2}}[LINX(\mathrm{N_2}-\mathrm{C_1}-\mathrm{N_1}-\mathrm{C_2})+LINX(\mathrm{N_3}-\mathrm{C_2}-\mathrm{N_1}-\mathrm{C_1})] \\
S_5(b_2) &= \frac {1}{\sqrt{2}}[(\mathrm{N_1}-\mathrm{C_1})-(\mathrm{N_1}-\mathrm{C_2})]\\
S_6(b_2) &= \frac {1}{\sqrt{2}}[(\mathrm{C_1}-\mathrm{N_2})-(\mathrm{C_2}-\mathrm{N_3})]\\
S_7(b_2) &= \frac {1}{\sqrt{2}}[LINX(\mathrm{N_2}-\mathrm{C_1}-\mathrm{N_1}-\mathrm{C_2})-LINX(\mathrm{N_3}-\mathrm{C_2}-\mathrm{N_1}-\mathrm{C_1})] \\
S_8(b_1) &= \frac {1}{\sqrt{2}}[LINY(\mathrm{N_2}-\mathrm{C_1}-\mathrm{N_1}-\mathrm{C_2})-LINY(\mathrm{N_3}-\mathrm{C_2}-\mathrm{N_1}-\mathrm{C_1})] \\
S_9(a_2) &= \frac {1}{\sqrt{2}}[LINY(\mathrm{N_2}-\mathrm{C_1}-\mathrm{N_1}-\mathrm{C_2})+LINY(\mathrm{N_3}-\mathrm{C_2}-\mathrm{N_1}-\mathrm{C_1})]
\end{align}
Similar internal coordinates have been used in cyclopropenylidene and its
silated cousin $c$-SiC$_2$H$_2$ \citep{Fortenberry18C3H2, Fortenberry18SiC2H2}.
The LINX and LINY coordinates have been defined previously by
\cite{Fortenberry12HOCO+}.

The CcCR QFF requires seven energy computations at each point.  The
CCSD(T)/aug-cc-pVTZ, -pVQZ, and -pV5Z energies are extrapolated to the complete
basis set (CBS) limit via a 3-point formula \citep{Martin96}.  The differences
in the CCSD(T)/MT energies are then added to this CBS energy, as are
differences in the CCSD(T)/cc-pVTZ-DK energies \citep{Douglas74, deJong01} with
and without relativity included.  This molecule represents the most basis
functions ever utilized for a full CcCR QFF computation.  Again, the F12-TZ QFF
only requires the CCSD(T)-F12/cc-pVTZ-F12 energies to be computed reducing the
computation time of the QFF from roughly two months in the CcCR approach to two
days with F12-TZ.  The CcCR center-of-mass dipole moment is computed at the
equilibrium geometry defined below with the CCSD(T)/aug-cc-pV5Z level of
theory.  The F12-TZ dipole moment is computed at its optimized geometry.  All
of the electronic structure computations described thus far employ the MOLPRO
2015.1 electronic structure program \citep{MOLPRO, MOLPRO-WIREs}.  The
double-harmonic intensities of the vibrational frequencies are computed with
MP2/6-31+G(d) in the Gaussian09 program \citep{MP2, Hehre72, g09} and are known
to closely mirror experimental and higher-order quantum chemical values
\citep{Fortenberry14LLL, Yu15NNHNN+, Finney16}.

Once the points are generated and the energies compiled, the total energies are
zeroed to the minimum energy, producing relative energies, which reduces the possible
numerical noise in the computation.  Then, a least squares fit procedure with a
sum of squared residuals on the order of $10^{-17}$ a.u.$^2$ for CcCR and
$10^{-18}$ a.u.$^2$ for F12-TZ is refit to the new minimum producing the
necessary force constants for the actual Taylor series definition of the
potential.  Transformation of the force constants into Cartesian derivatives
via the Intder program \citep{intder} allows for more flexible usage of the
potential within the Spectro program \citep{spectro91}.  Spectro utilizes
rotational and vibrational perturbation theory at second order (VPT2)
\citep{Mills72, Watson77, Papousek82} to produce the observable spectroscopic
values.  NCNCN$^-$ exhibits a $2\nu_6=2\nu_7=2\nu_8=\nu_4$ Fermi resonance
polyad, a $\nu_4+\nu_3=\nu_2$ type-2 Fermi resonance, and a
$\nu_8+\nu_4=\nu_8+\nu_5=\nu_3$ Fermi resonance polyad.  Additionally, four
Coriolis resonances are also treated in this system: $\nu_6$/$\nu_5$ B$-$type,
$\nu_7$/$\nu_5$ A$-$type, $\nu_8$/$\nu_6$ A$-$type, and $\nu_8$/$\nu_7$
B$-$type.  The inclusion of these resonances has been shown to further improve
the accuracy of QFF VPT2-type approaches markedly \citep{Maltseva15}.

\section{Results}

\subsection{Structure and Rotational Spectroscopic Constants}

\renewcommand{\baselinestretch}{1}
\begingroup
\begin{table}

\caption{The CcCR and F12/TZ Equilibrium and Vibrationally-Averaged
($R_{\alpha}$) Structural and Rotational Spectroscopic Constants.}

\label{Struct}

\centering
\scriptsize

\begin{tabular}{l | l | c c}
\hline
 & Units & CcCR & F12-TZ \\
\hline
r$_e$(N$_1-$C$_{1/2}$)        & \AA       & 1.31467 & 1.31834  \\
r$_e$(C$_{1/2}\equiv$N$_{2/3}$)& \AA      & 1.17199 & 1.17504  \\
$\angle$$_e$(C$_1-$N$_1-$C$_2$) &$^{\circ}$ & 118.82 & 118.50  \\
$\angle$$_e$(N$_1-$C$_{1/2}-$N$_{2/3}$)&$^{\circ}$ & 174.12 & 174.01  \\
A$_e$                         & MHz       & 39786.3 & 39290.3  \\
B$_e$                         & MHz       &  3047.3 &  3038.1  \\
C$_e$                         & MHz       &  2830.5 &  2820.0  \\
D$_J$                         & kHz       &  1.290  &  1.282   \\
D$_{JK}$                      & MHz       & -0.152  & -0.148   \\
D$_K$                         & MHz       &  6.009  &  5.778   \\
d$_1$                         & kHz       & -0.294  & -0.294   \\
d$_2$                         & kHz       & -0.005  & -0.005   \\
H$_J$                         & mHz       &  3.753  &  3.738   \\
H$_{JK}$                      &  Hz       & -0.083  & -0.089   \\
H$_{KJ}$                      &  Hz       &-60.234  &-56.845   \\
H$_K$                         & kHz       &  2.952  &  2.754   \\
h$_1$                         & mHz       &  1.415  &  1.412   \\
h$_2$                         & mHz       &  0.061  &  0.062   \\
h$_3$                         & mHz       &  0.023  &  0.023   \\
$\mu$                         & D         &  1.03   & 1.02\\
\hline
r$_0$(N$_1-$C$_{1/2}$)        & \AA       & 1.31868 & 1.32234  \\
r$_0$(C$_{1/2}\equiv$N$_{2/3}$)& \AA      & 1.17264 & 1.17560  \\
$\angle$$_0$(C$_1-$N$_1-$C$_2$) &$^{\circ}$ & 118.78 & 118.46 \\
$\angle$$_0$(N$_1-$C$_{1/2}-$N$_{2/3}$)&$^{\circ}$ & 174.04 & 173.94  \\
A$_0$                         & MHz       & 40289.7 & 39776.3 \\
B$_0$                         & MHz       & 3037.4  & 3028.5  \\
C$_0$                         & MHz       & 2819.0  & 2808.8  \\
A$_1$                         & MHz       & 40228.9 & 39713.8 \\
B$_1$                         & MHz       & 3025.6  & 3016.8  \\
C$_1$                         & MHz       & 2808.6  & 2798.4  \\
A$_2$                         & MHz       & 39938.8 & 39435.6 \\
B$_2$                         & MHz       & 3029.7  & 3020.7  \\
C$_2$                         & MHz       & 2810.6  & 2800.4  \\
A$_3$                         & MHz       & 38738.7 & 38268.8 \\
B$_3$                         & MHz       & 3052.5  & 3043.3  \\
C$_3$                         & MHz       & 2824.6  & 2814.3  \\
A$_4$                         & MHz       & 41207.2 & 40660.4 \\
B$_4$                         & MHz       & 3017.2  & 3008.6  \\
C$_4$                         & MHz       & 2804.2  & 2794.2  \\
A$_5$                         & MHz       & 41093.4 & 40558.2 \\
B$_5$                         & MHz       & 3026.5  & 3017.7  \\
C$_5$                         & MHz       & 2812.9  & 2802.8  \\
A$_6$                         & MHz       & 40102.4 & 39604.7 \\
B$_6$                         & MHz       & 3043.0  & 3034.2  \\
C$_6$                         & MHz       & 2824.0  & 2814.0  \\
A$_7$                         & MHz       & 40769.0 & 40242.9 \\
B$_7$                         & MHz       & 3029.4  & 3021.0  \\
C$_7$                         & MHz       & 2816.4  & 2806.6  \\
A$_8$                         & MHz       & 40112.1 & 39606.5 \\
B$_8$                         & MHz       & 3045.4  & 3036.4  \\
C$_8$                         & MHz       & 2823.7  & 2813.5  \\
A$_9$                         & MHz       & 41424.5 & 40867.9 \\
B$_9$                         & MHz       & 3047.4  & 3038.7  \\
C$_9$                         & MHz       & 2823.2  & 2813.0  \\
\hline                       

\end{tabular}

\end{table}
\endgroup 
\renewcommand{\baselinestretch}{2}

NCNCN$^-$ is a near-prolate molecule of likely high detectability.  -CN groups
have been some of the most observed functional groups to date in interstellar
molecules \citep{McGuire18Census}, and the anion of interest here has
two.  According to Table \ref{Struct}, the CcCR C$\equiv$N bond lengths of
1.17264 \AA\ are nearly exactly the value expected for such bonding, like in the
1.172 \AA\ bond length in the cyano radical \citep{Huber}, and they are very
nearly the same for F12-TZ.  Additionally, the F$_{33}$ force constant in Table
\ref{fcs} is rather large at 16.514 mdyn/\AA$^2$ indicating very strong bonding
typical of a cyano group.  In comparison, Fortenberry et al. 2013 had calculated an F$_{33}$ force constant of 15.589 mdyn/\AA$^2$ for the cyanomethyl anion CH$_2$CN$^-$.   The N$_1-$C$_{1/2}$ bonding is also strong at 7.036
mdyn/\AA$^2$, but this is closer to the values expected of a single bond as the
Lewis structure of this molecule would indicate.  The triple bond also produces
a near-linearity in the $\angle$(N$_1-$C$_{1/2}-$N$_{2/3}$) bond angles which
are 174.04$^{\circ}$ in line with previous work.  The rest of the F12-TZ force
constants necessary for the Spectro computations are in Table \ref{fcs}.  

\renewcommand{\baselinestretch}{1}
\begingroup
\setlength{\tabcolsep}{10pt}
\begin{table}

\caption{The F12-TZ Force Constants (in mdyn/\AA$^n$$\cdot$rad$^m$) with the
Indices Corresponding to the Coordinate Numbers Given in the Text.}

\label{fcs}

\centering
\footnotesize

\begin{tabular}{||l |r| |l |r| |l |r| |l |r| |l |r|}

\hline
F$_{11}$ &  1.569243 & F$_{661}$ &   0.0091 & F$_{4311}$ &    0.48 & F$_{6643}$ &   -0.11 & F$_{8821}$ &   -0.17 \\
F$_{21}$ & -0.401374 & F$_{662}$ &  -0.2649 & F$_{4321}$ &   -0.51 & F$_{6644}$ &   -0.01 & F$_{8822}$ &    0.28 \\
F$_{22}$ &  7.035627 & F$_{663}$ & -77.8702 & F$_{4322}$ &    0.90 & F$_{6655}$ &    2.45 & F$_{8831}$ &   -0.15 \\
F$_{31}$ & -0.141918 & F$_{664}$ &  -0.1133 & F$_{4331}$ &    0.21 & F$_{6665}$ &   -1.72 & F$_{8832}$ &    1.05 \\
F$_{32}$ &  0.689322 & F$_{751}$ &   0.2937 & F$_{4332}$ &   -0.25 & F$_{6666}$ &  300.54 & F$_{8833}$ &    0.18 \\
F$_{33}$ & 16.514247 & F$_{752}$ &  -0.4648 & F$_{4333}$ &   -0.03 & F$_{7511}$ &    0.09 & F$_{8841}$ &   -0.02 \\
F$_{41}$ &  0.110356 & F$_{753}$ &   0.0150 & F$_{4411}$ &    0.65 & F$_{7521}$ &   -0.72 & F$_{8842}$ &    0.14 \\
F$_{42}$ &  0.089479 & F$_{754}$ &  -0.6003 & F$_{4421}$ &   -0.70 & F$_{7522}$ &    1.66 & F$_{8843}$ &    0.08 \\
F$_{43}$ &  0.065200 & F$_{761}$ &  -0.0776 & F$_{4422}$ &    1.04 & F$_{7531}$ &    0.12 & F$_{8844}$ &    0.49 \\
F$_{44}$ &  0.561455 & F$_{762}$ &  -0.0910 & F$_{4431}$ &   -0.02 & F$_{7532}$ &   -0.11 & F$_{8855}$ &    0.37 \\
F$_{55}$ &  8.015981 & F$_{763}$ &  -0.0639 & F$_{4432}$ &    0.73 & F$_{7533}$ &    0.11 & F$_{8865}$ &    0.64 \\
F$_{65}$ &  0.954895 & F$_{764}$ &  -0.7086 & F$_{4433}$ &    0.12 & F$_{7541}$ &   -0.11 & F$_{8866}$ &    0.10 \\
F$_{66}$ & 16.160572 & F$_{771}$ &   0.0700 & F$_{4441}$ &    0.12 & F$_{7542}$ &    0.98 & F$_{8875}$ &   -0.10 \\
F$_{75}$ &  0.028367 & F$_{772}$ &  -0.7099 & F$_{4442}$ &    0.16 & F$_{7543}$ &    0.56 & F$_{8876}$ &    0.06 \\
F$_{76}$ &  0.059044 & F$_{773}$ &  -0.8253 & F$_{4443}$ &    0.17 & F$_{7544}$ &    0.10 & F$_{8877}$ &    0.54 \\
F$_{77}$ &  0.559994 & F$_{774}$ &  -0.2081 & F$_{4444}$ &    1.50 & F$_{7555}$ &   -0.24 & F$_{8888}$ &    1.66 \\
F$_{88}$ &  0.584796 & F$_{881}$ &  -0.0435 & F$_{5511}$ &    2.40 & F$_{7611}$ &   -0.09 & F$_{9851}$ &    0.08 \\
F$_{99}$ &  0.636805 & F$_{882}$ &  -0.5506 & F$_{5521}$ &    0.15 & F$_{7621}$ &   -0.07 & F$_{9852}$ &    0.79 \\
F$_{111}$ &   1.0720 & F$_{883}$ &  -0.7621 & F$_{5522}$ &  131.56 & F$_{7622}$ &    0.41 & F$_{9853}$ &    0.56 \\
F$_{211}$ &  -1.7283 & F$_{884}$ &  -0.0571 & F$_{5531}$ &   -0.79 & F$_{7631}$ &   -0.01 & F$_{9854}$ &    0.13 \\
F$_{221}$ &  -0.8397 & F$_{985}$ &  -0.8259 & F$_{5532}$ &    3.72 & F$_{7632}$ &   -0.01 & F$_{9861}$ &   -0.18 \\
F$_{222}$ & -31.2465 & F$_{986}$ &  -0.5932 & F$_{5533}$ &    0.51 & F$_{7633}$ &   -0.10 & F$_{9862}$ &    0.61 \\
F$_{311}$ &  -0.8862 & F$_{987}$ &  -0.0609 & F$_{5541}$ &    0.17 & F$_{7641}$ &    0.12 & F$_{9863}$ &   -0.02 \\
F$_{321}$ &   1.3618 & F$_{991}$ &  -0.0345 & F$_{5542}$ &   -0.37 & F$_{7642}$ &    0.54 & F$_{9864}$ &    0.07 \\
F$_{322}$ &  -3.0329 & F$_{992}$ &  -0.5765 & F$_{5543}$ &    0.09 & F$_{7643}$ &   -0.04 & F$_{9871}$ &   -0.02 \\
F$_{331}$ &  -0.1273 & F$_{993}$ &  -0.8648 & F$_{5544}$ &   -0.04 & F$_{7644}$ &    0.09 & F$_{9872}$ &    0.13 \\
F$_{332}$ &  -0.7153 & F$_{994}$ &  -0.1138 & F$_{5555}$ &  117.81 & F$_{7655}$ &   -0.15 & F$_{9873}$ &    0.05 \\
F$_{333}$ & -78.3673 & F$_{1111}$ &    5.27 & F$_{6511}$ &    0.93 & F$_{7665}$ &   -0.01 & F$_{9874}$ &    0.47 \\
F$_{411}$ &  -0.3095 & F$_{2111}$ &   -2.40 & F$_{6521}$ &   -0.84 & F$_{7666}$ &   -0.18 & F$_{9911}$ &   -0.02 \\
F$_{421}$ &   0.4662 & F$_{2211}$ &   -1.74 & F$_{6522}$ &    3.14 & F$_{7711}$ &    0.74 & F$_{9921}$ &    0.00 \\
F$_{422}$ &  -1.0531 & F$_{2221}$ &   10.99 & F$_{6531}$ &   -0.42 & F$_{7721}$ &   -0.97 & F$_{9922}$ &   -0.10 \\
F$_{431}$ &  -0.1847 & F$_{2222}$ &  102.54 & F$_{6532}$ &    2.88 & F$_{7722}$ &    1.49 & F$_{9931}$ &   -0.22 \\
F$_{432}$ &  -0.0670 & F$_{3111}$ &   -0.51 & F$_{6533}$ &   -1.42 & F$_{7731}$ &   -0.21 & F$_{9932}$ &    1.19 \\
F$_{433}$ &  -0.0812 & F$_{3211}$ &    2.97 & F$_{6541}$ &   -0.13 & F$_{7732}$ &    1.16 & F$_{9933}$ &    0.13 \\
F$_{441}$ &  -0.0244 & F$_{3221}$ &   -7.56 & F$_{6542}$ &    0.07 & F$_{7733}$ &    0.10 & F$_{9941}$ &    0.07 \\
F$_{442}$ &  -0.5255 & F$_{3222}$ &   15.94 & F$_{6543}$ &   -0.09 & F$_{7741}$ &   -0.07 & F$_{9942}$ &    0.11 \\
F$_{443}$ &  -0.8250 & F$_{3311}$ &   -0.73 & F$_{6544}$ &    0.74 & F$_{7742}$ &    0.40 & F$_{9943}$ &    0.06 \\
F$_{444}$ &  -0.1967 & F$_{3321}$ &    0.44 & F$_{6555}$ &    3.21 & F$_{7743}$ &    0.20 & F$_{9944}$ &    0.54 \\
F$_{551}$ &  -1.0580 & F$_{3322}$ &    1.23 & F$_{6611}$ &    0.05 & F$_{7744}$ &    1.57 & F$_{9955}$ &    0.42 \\
F$_{552}$ & -35.9427 & F$_{3331}$ &    0.11 & F$_{6621}$ &   -1.03 & F$_{7755}$ &    0.16 & F$_{9965}$ &    0.82 \\
F$_{553}$ &  -2.4013 & F$_{3332}$ &    0.19 & F$_{6622}$ &    3.02 & F$_{7765}$ &    0.69 & F$_{9966}$ &   -0.01 \\
F$_{554}$ &  -0.0448 & F$_{3333}$ &  299.39 & F$_{6631}$ &    0.27 & F$_{7766}$ &    0.04 & F$_{9975}$ &    0.16 \\
F$_{651}$ &   0.2463 & F$_{4111}$ &   -3.18 & F$_{6632}$ &   -1.49 & F$_{7775}$ &    0.18 & F$_{9976}$ &    0.07 \\
F$_{652}$ &  -2.1510 & F$_{4211}$ &    4.07 & F$_{6633}$ &  299.62 & F$_{7776}$ &    0.10 & F$_{9977}$ &    0.58 \\
F$_{653}$ &  -0.6691 & F$_{4221}$ &   -4.80 & F$_{6641}$ &   -0.28 & F$_{7777}$ &    1.67 & F$_{9988}$ &    1.75 \\
F$_{654}$ &   0.0869 & F$_{4222}$ &    4.80 & F$_{6642}$ &    0.22 & F$_{8811}$ &    0.03 & F$_{9999}$ &    1.94 \\
\hline                       
\end{tabular}
\end{table}
\endgroup 
\renewcommand{\baselinestretch}{2}

The structure of this molecule from such near-linearities lends itself to being
near-prolate as the rotational constants in Table \ref{Struct} indicate.  The B
and C rotational constants are a factor of more than 13 less than the A
constant indicating that the K-branching would typically be small and
potentially non-existent at lower resolutions.  
However, the 1.03 D dipole moment is aligned along the C rotational axis and therefore the analysis of any future observed rotational spectrum of this molecule will require all three constants to fit its Hamiltonian.  This molecule has the potential to be observable with ground-based radio telescopes.
Even though the F12-TZ rotational constants are known to not be as good as the CcCR values unlike the vibrational frequencies \citep{Agbaglo19b} which are discussed below, the B and C constants vary by around only 10 MHz between the QFFs.  
The quartic and sextic distortion constants (D and H) as well as the vibrationally-excited rotational constants (numbered from Table \ref{freqs})
are also given in Table \ref{Struct}.  These distortions are relatively small
due to the rigidity of the molecule.  The vibrationally-excited B and C
constants do not vary greatly with vibrational excitation, but the A constants
oscillate by as much as 3000 MHz from mode to mode.

\subsection{Fundamental Vibrational Frequencies}

\renewcommand{\baselinestretch}{1}
\begingroup
\begin{table}

\caption{The CcCR and F12-TZ QFF VPT2 Harmonic and Anharmonic Vibrational
Frequencies in cm$^{-1}$ with Intensities in km/mol.$^a$}

\label{freqs}

\centering

\begin{tabular}{l | c | c c | c}
\hline
 & Description & CcCR & F12-TZ & Intensity\\
\hline
$\omega_1$ ($a_1$) & C symm.~stretch     & 2232.7 & 2224.5 & (97)  \\
$\omega_2$ ($b_2$) & C antisymm.~stretch & 2205.4 & 2197.7 & (1124)\\
$\omega_3$ ($b_2$) & N antisymm.~stretch & 1322.9 & 1314.7 & (129) \\
$\omega_4$ ($a_1$) & N symm.~stretch     &  920.8 &  918.3 & (17)  \\
$\omega_5$ ($b_2$) & symm.~N$-$C$-$N bend&  670.8 &  668.2 & (3)   \\
$\omega_6$ ($a_2$) & antisymm.~OPB       &  553.5 &  549.4 &       \\
$\omega_7$ ($b_2$) & antisymm.~N$-$C$-$N &  539.1 &  534.2 & (5)   \\
$\omega_8$ ($b_1$) & symm.~OPB           &  514.2 &  509.9 & (17)  \\
$\omega_9$ ($a_1$) & C$-$N$-$C bend      &  165.9 &  165.8 & (5)   \\
\hline
$\nu_1$ ($a_1$) & C symm.~stretch      & 2190.7 & 2185.5  \\
$\nu_2$ ($b_2$) & C antisymm.~stretch  & 2130.9 & 2123.8  \\
$\nu_3$ ($b_2$) & N antisymm.~stretch  & 1302.3 & 1293.7  \\
$\nu_4$ ($a_1$) & N symm.~stretch      &  901.3 &  894.1  \\
$\nu_5$ ($b_2$) & symm.~N$-$C$-$N bend &  661.0 &  660.5  \\
$\nu_6$ ($a_2$) & antisymm.~OPB        &  549.0 &  544.9  \\
$\nu_7$ ($b_2$) & antisymm.~N$-$C$-$N  &  532.4 &  527.9  \\
$\nu_8$ ($b_1$) & symm.~OPB            &  510.3 &  505.4  \\
$\nu_9$ ($a_1$) & C$-$N$-$C bend       &  162.6 &  164.0  \\
ZPVE            &                      & 4535.9 & 4515.9 \\
$2\nu_1$        &                      & 4369.1 & 4358.8 \\
$\nu_1+\nu_2$   &                      & 4329.4 & 4320.0 \\
$2\nu_2$        &                      & 4312.1 & 4303.4 \\
$\nu_1+\nu_3$   &                      & 3479.7 & 3465.8 \\
$\nu_2+\nu_3$   &                      & 3450.0 & 3436.9 \\
$2\nu_3$        &                      & 2577.7 & 2560.5 \\
\hline

\end{tabular}
\\$^a$Double-harmonic MP2/6-31+G(d) intensities.\\

\end{table}
\endgroup
\renewcommand{\baselinestretch}{2}

NCNCN$^-$ has an exceptionally bright $\nu_2$ fundamental vibrational frequency
at \\ 2130.9 cm$^{-1}$ or 4.693 $\mu$m.  The intensity is predicted to be on the
order of that expected for exotic species like proton-bound complexes
\citep{Fortenberry15OCHCO+, Yu15NNHNN+, McDonald16, Stephan17} and azide anions
\citep{Kelly18}.  The proton-bound complexes shuttle nearly all of the charge
and nearly none of the mass back and forth creating a huge induced dipole shift
in the hydrogen atom's antisymmetric stretch.  Here in NCNCN$^-$ the effect is
slightly different in that the bond order is decreasing in one direction of the
individual carbon atom's movement and increasing in the other, and there are
two such motions happening conversely on opposite sides of the molecule.
Consequently, a significant amount of electronic charge is moving in one
direction across the entire molecule in much the same way but to a lesser
extent than what happens in a plasmon.  Additionally, this region of the
infrared is relatively quiet in most astronomical spectra with few molecules
exhibiting much motion due to the gulf in mass between hydrogen and any
astrophysically-meaningful heavier atoms.  Granted, there are some lines and
even a few bright ones.  Hence, this molecule could serve to be a possible
carrier in this region.

Additionally, the $\nu_1$ fundamental at 2190.7 cm$^{-1}$ or 4.565 $\mu$m is
also notable but not nearly as bright.  The same bond order increases and
decreases are still happening due again to the motion of the internal carbon
atoms, but now they are in concert only pushing the charge toward the central
N$_1$ atom.  Regardless, these two vibrational frequencies are probably convolved due to their close frequencies and wavelengths with $\nu_2$
dominating to a significant extent.  Even so, they both should be clear markers
for observations with NIRSpec on the upcoming James Webb Space Telescope
(JWST).

The $\nu_3$ fundamental at 1302.3 cm$^{-1}$ or 7.679 $\mu$m is predicted to be
of the same intensity magnitude as the $\nu_1$ but much more reddened.  With
an antisymmetric stretch of the terminal nitrogen atoms, the amount of charge
movement is not as great as $\nu_2$ since each nitrogen atom is only bonded to
one carbon atom.  Even so, this fundamental will likely not be observable
astronomically as it sits squarely in a region almost certainly dominated by
polycyclic aromatic hydrocarbons (PAHs).  Hence, this frequency could assist in
laboratory analysis but likely not in astrophysical classification.

The rest of the fundamental vibrational frequencies are relatively dim and will
be further clouded by the myriad of other infrared features from various other
molecular systems such as PAHs or silicates much more active below 1000
cm$^{-1}$, or 10 $\mu$m.  However, the overtones and combination bands of the
bright fundamentals could be observed in near-IR, especially with JWST and SOFIA (the Stratospheric Observatory for Infrared Astronomy).
The best candidate is the $2\nu_2$ at 4312.1 cm$^{-1}$ or 2.319 $\mu$m due to
the large intensity of the fundamental.  The two combination bands of $\nu_2$
with the other bright fundamentals ($\nu_1+\nu_2$ and $\nu_2+\nu_3$) could also
possibly be observed at 4329.4 cm$^{-1}$ (2.304 $\mu$m) and 3450.0 cm$^{-1}$
(2.899 $\mu$m), respectively.  However, the $\nu_1+\nu_2$ would probably be
conjoined with $2\nu_2$ much more than even the $\nu_1$ and $\nu_2$ fundamentals would, making $\nu_1+\nu_2$ a shoulder on the taller $2\nu_2$ peak at best.  The
$\nu_2+\nu_3$ overtone could also be observed but likely only in regions where
this anion is highly abundant.

In a general sense for this molecule, none of the fundamentals are highly
anharmonic as would be expected for a molecule with no hydrogen atoms present.
The $\nu_1$ fundamental has an anharmonicity on the order of 40 cm$^{-1}$ which
still requires some type of anharmonic computation in order to begin to match
laboratory or interstellar observations.  The $\nu_2$ fundamental has an
anharmonicity reduction of slightly less at around 35 cm$^{-1}$.  Hence, static
scaling factors on harmonic computations could not reproduce the likely
reliability of the frequency values here.

Additionally, the CcCR and F12-TZ QFF results match very closely for the
vibrational frequencies, especially in the anharmonic values.  The largest
difference is for $\nu_3$ at 8.6 cm$^{-1}$, and most are less than 5.0
cm$^{-1}$ in line with what has previously been shown for closed-shell
molecules even anions \citep{Agbaglo19b}.  Hence, the values produced here give
no indication of any irregularities and should assist in further spectral
characterization of this markedly stable anion.

\section{Conclusions}

Using a quantum chemical analysis, we have demonstrated that C$_2$N$_3$$^-$ displays a notable $\nu _2$  fundamental vibrational frequency at 2130.9 cm$^{-1}$ (4.693 $\mu m$) carried by an antisymmetric stretch of the carbon atoms, while their symmetric $\nu _1$ stretch vibration occurs at 2190.7 cm$^{-1}$ (4.565 $\mu$m). The two terminal nitrogen atoms carry the antisymmetric stretch fundamental $\nu _3$ at 1302.3 cm$^{-1}$ (7.679 $\mu$m), a region rich in polycyclic aromatic hydrocarbons (PAHs). Near-IR combination and overtone bands are present below 2.9 $\mu$m and could also be used for astrochemical observations, such as the $2\nu _2$ at 4312.1 (2.319 $\mu$m). The F12-TZ approach is once more shown to be as reliable as the CcCR protocol indicating that it will continue to be useful for future work in this area. Overall, these quantum chemical calculations should aid in the search and detection of this remarkably stable anion in astrochemically-relevant environments such as in the atmospheres of carbon-rich stars or Saturn's moon Titan, particularly in the era of the upcoming James Webb Space Telescope.

\section{Acknowledgements}

RCF acknowledges support from NASA grant NNX17AH15G, NSF grant OIA-1757220, and
start-up funds provided by the University of Mississippi. DD expresses gratitude towards the NASA SMD (NNH17ZDA001N-CDAP) for funding, as well as Dr. Partha Bera for fruitful discussions. ESO acknowledges support from NASA grants NNH17ZDA001N-CDAP and NNH17ZDA001N-SSW.

\bibliographystyle{apj}
\bibliography{fullrefsJuly10}

\end{document}